\RequirePackage{fix-cm}
\documentclass[twocolumn,epjc3]{svjour3}  
\smartqed  
\RequirePackage{color, graphicx}
\RequirePackage{mathptmx}      
\RequirePackage{latexsym}
\RequirePackage{amsmath}
\RequirePackage[numbers,sort&compress]{natbib}
\RequirePackage[colorlinks,citecolor=blue,urlcolor=blue,linkcolor=blue]{hyperref}
\journalname{Eur. Phys. J. C}
\begin{document}

\title{Searching for degenerate Higgs bosons
}
\subtitle{A profile likelihood ratio method to test for mass-degenerate states in the presence of incomplete data and uncertainties}


\author{
        Andr\'{e} David\thanksref{emailAD,CERN}
        \and
        Jaana Heikkil\"{a}\thanksref{emailJH,CERN,Helsinki}
        \and
        Giovanni Petrucciani\thanksref{emailGP,CERN}
}

\thankstext{emailAD}{e-mail: Andre.David@cern.ch}
\thankstext{emailJH}{e-mail: Jaana.Kristiina.Heikkilae@cern.ch}
\thankstext{emailGP}{e-mail: Giovanni.Petrucciani@cern.ch}


\institute{PH Department, CERN, Switzerland \label{CERN}
           \and
           Helsinki Institute of Physics, Finland\label{Helsinki}
}

\date{Received: date / Accepted: date}

\maketitle

\begin{abstract}
Using the likelihood ratio test statistic, we present a method which can be employed to test the hypothesis of a single Higgs boson using the matrix of measured signal strengths.
This method can be applied in the presence of incomplete data and takes into account uncertainties on the measurements.
The p-value against the hypothesis of a single Higgs boson is defined from the expected distribution of the test statistic, generated using pseudo-experiments.
The applicability of the likelihood-based test is demonstrated using numerical examples with uncertainties and missing matrix elements.

\end{abstract}

\section{Introduction}
\label{intro}
A new resonance, consistent with the standard model (SM) Higgs boson and with a mass of approximately 125~GeV, has been observed by ATLAS and CMS collaborations at the Large Hadron Collider (LHC) \cite{AtlasObservation2012,CMSObservation2012,CMSLong2013}.
For each production mode and decay mode of the boson, the signal strength can be been defined in terms of the observed production cross section and branching ratio relative to the SM expectation as:
\begin{equation}\label{mudef}
\mu_{i,j}=\frac{(\sigma_{i}\cdot \mathcal{B}_{j})_\mathrm{obs}}{(\sigma_{i}\cdot \mathcal{B}_{j})_\mathrm{SM}},
\end{equation}
where the indexes $i$ and $j$ stand for the different production modes and decay modes, respectively.
The available experimental information can be arranged in a matrix, $\mathcal{M}$, of signal strengths:
\begin{equation}\label{5x4}
\begin{array}{r|ccccc}
& \mathrm{H}\rightarrow \gamma\gamma & \mathrm{H}\rightarrow \mathrm{WW} & \mathrm{H}\rightarrow \mathrm{ZZ} & \mathrm{H}\rightarrow \tau\tau & \mathrm{H}\rightarrow \mathrm{bb}\\
 \hline
\mathrm{ggH} & \mu_{\mathrm{ggH},\gamma\gamma} & \mu_{\mathrm{ggH,WW}} & \mu_{\mathrm{ggH,ZZ}} & \mu_{\mathrm{ggH},\tau\tau} & \mu_{\mathrm{ggH,bb}}\\[3pt]
\mathrm{VBF} & \mu_{\mathrm{VBF},\gamma\gamma} & \mu_{\mathrm{VBF,WW}} & \mu_{\mathrm{VBF,ZZ}} & \mu_{\mathrm{VBF},\tau\tau} & \mu_{\mathrm{VBF,bb}} \\[3pt]
\mathrm{VH} & \mu_{\mathrm{VH},\gamma\gamma} & \mu_{\mathrm{VH,WW}} & \mu_{\mathrm{VH,ZZ}} & \mu_{\mathrm{VH},\tau\tau} & \mu_{\mathrm{VH,bb}} \\[3pt]
\mathrm{ttH} & \mu_{\mathrm{ttH},\gamma\gamma} & \mu_{\mathrm{ttH,WW}} & \mu_{\mathrm{ttH,ZZ}} & \mu_{\mathrm{ttH},\tau\tau} & \mu_{\mathrm{ttH,bb}}
\end{array}
\end{equation}

The rows of $\mathcal{M}$ represent the four main production modes: gluon fusion (ggH), vector boson fusion (VBF), associated production with a Z or W boson (VH), and associated production with a top quark pair (ttH).
The columns represent five decay modes\footnote{We imply charge conjugation throughout; $\mathrm{bb}$ stands for $\mathrm{b\bar{b}}$, $\tau\tau$ stands for $\tau^{+}\tau^{-}$, etc.}: $\gamma\gamma$, $\mathrm{WW}$, $\mathrm{ZZ}$, $\tau\tau$ and $\mathrm{bb}$.

It has been suggested that there may be more than one state nearly-degenerate in mass, which could couple differently to the SM particles and produce the observed signal.
If more than one state is present, it may be possible to use the rank of $\mathcal{M}$ to test for multiple states without having to directly resolve the resonant peak(s), an approach limited by the experimental mass resolution to few GeV~\cite{Grossman,ATLASMassLegacyRun1,CMSHggLegacyRun1}.
To that end, there are two main challenges that need to be overcome in order to make use of all the experimental information encoded in $\mathcal{M}$:
\begin{enumerate}
\item Some $\mu_{i,j}$ have not yet been measured, such as $\mu_{\mathrm{ttH,ZZ}}$, and are therefore missing, raising the issue of incomplete data.
\item The precision to which different $\mu_{i,j}$ have been measured varies greatly, with the smaller uncertainties usually being related to the processes with the largest production cross-sections.
\end{enumerate}

Inspired by the work presented in Ref.~\cite{Grossman}, we noticed that while their approach tackled the uncertainty in the measurements, it was not able to take into account the missing data aspect.
The method presented here addresses the question of whether the experimental observations are compatible with a single Higgs boson using the full information encoded in $\mathcal{M}$, i.e.\ coping with missing $\mu_{i,j}$ and taking into account the uncertainties on the $\mu_{i,j}$.

\section{The $2\times2$ case}
\label{synExample}
Consider a $2\times2$ sub-matrix of $\mathcal{M}$, namely the one with the ggH and VBF  production modes and the  $\gamma\gamma$ and WW decay channels:
\begin{equation}\label{2x2ggWW}
\begin{array}{r|cc}
& \mathrm{H}\rightarrow\gamma\gamma & \mathrm{H}\rightarrow
 \mathrm{WW} \\
 \hline
 \mathrm{ggH} & \mu_{\mathrm{ggH}, \gamma\gamma}   & \mu_{\mathrm{ggH, WW}} \\[3pt]
 \mathrm{VBF} & \mu_{\mathrm{VBF}, \gamma\gamma}   & \mu_{\mathrm{VBF, WW}} 
\end{array}
\end{equation}

Ignoring the uncertainties on the $\mu_{i,j}$ elements, if there is only one Higgs boson the determinant of this $2\times2$ matrix is zero and consequently the matrix has rank 1.

However, if there are two particles involved, each signal strength in the matrix can be written as the sum of two terms:
\begin{equation}\label{sum}
\mu_{i,j}=\mu_{i,j}^{(1)}+\mu_{i,j}^{(2)},
\end{equation}
where the superscripts 1 and 2 concern each of the two states.
When replacing all four matrix elements with sums of form Eq.~(\ref{sum}), the determinant becomes, in general, not equal to zero and in that case the matrix rank will be 2.

Instead of the determinant, when studying the rank of a $2\times2$ matrix, a double ratio can be used without loss of generality:
\begin{equation}\label{rho}
\rho=\frac{\mu_{\mathrm{ggH}, \gamma\gamma} \ \mu_{\mathrm{VBF, WW}}}{\mu_{\mathrm{ggH, WW}} \ \mu_{\mathrm{VBF}, \gamma\gamma}}.
\end{equation}
This is equivalent to the double ratios introduced in Ref.~\cite{Gunion} and the expectation is that $\rho=1$ if there is only one state.
If there are two states, one has
\begin{equation}
\rho=\frac{(\mu_{\mathrm{ggH}, \gamma\gamma}^{(1)}+\mu_{\mathrm{ggH}, \gamma\gamma}^{(2)})\ (\mu_{\mathrm{VBF, WW}}^{(1)}+\mu_{\mathrm{VBF, WW}}^{(2)})}{(\mu_{\mathrm{ggH, WW}}^{(1)}+\mu_{\mathrm{ggH, WW}}^{(2)})\ (\mu_{\mathrm{VBF}, \gamma\gamma}^{(1)}+\mu_{\mathrm{VBF}, \gamma\gamma}^{(2)})},
\end{equation}
and $\rho$ can take a range of values, implying the rank of the matrix is 2.

The rank is a discrete quantity that can be exactly computed if the elements of the matrix are also exactly known.
If there are independent uncertainties on the matrix elements, the rank cannot be precisely determined.
Nevertheless, it is still possible to assess the statistical compatibility of the matrix with one of rank 1.
In previous works~\cite{Grossman,Gunion}, it was proposed to assess the statistical compatibility via quantities such as $\rho$ which can be assigned an uncertainty obtained by propagating the uncertainties on the matrix elements.

\section{The profile likelihood ratio test}
\label{likelihoodTest}

In order to evaluate the statistical compatibility of the matrix of observations having rank 1, a model of the matrix elements must be assumed.
Such a model defines the correlations between the different $\mu_{i,j}$, which are parameters of interest in the fit to data, but may also include all other parameters that affect the measurements, such as systematic uncertainties.
The latter can be treated as nuisance parameters using the profile likelihood method and their values are obtained from a fit to the data~\cite{Cowan}.

The likelihood is the probability to observe the data assuming that a given model is true.
Using the likelihood function $L(\mathrm{data}|\alpha, \beta)$, where $\alpha$ are parameters of interest and $\beta$ are nuisance parameters, we define the test statistic $q=q(\alpha)$ based on the profile likelihood ratio~\cite{NeymanPearson1,NeymanPearson2,PDG,Cowan,CowanBook,Kendall}:
\begin{equation}\label{testS}
q(\alpha)=-2\ln\frac{L(\mathrm{data}|\alpha, \hat{\hat{\beta}})}{L(\mathrm{data}|\hat{\alpha}, \hat{\beta})},
\end{equation}
where $\hat{\hat{\beta}}$ is the value of $\beta$ that maximizes the likelihood function for a particular value of $\alpha$, whereas $\hat{\alpha}$ and $\hat{\beta}$ are the values that maximize the likelihood function overall.

\subsection{Application to the $2\times2$ case}
\label{appTo2x2}
Let's consider again a $2\times2$ matrix and no nuisance parameters.
In this case, the test statistic $q(\rho)$ can be constructed for the double ratio $\rho$ from Eq.~\eqref{rho}:
\begin{equation}\label{qrho}
q(\rho)=-2\ln\frac{L(\mathrm{data}|\rho)}{L(\mathrm{data}|\hat{\rho})}.
\end{equation}
When testing for compatibility with rank 1, the relevant quantity is $q(\rho=1)$,
where the numerator is the likelihood to observe the data assuming a rank 1 matrix, whereas the denominator is the likelihood to observe the data assuming the most general $2\times2$ matrix.

\subsection{Application to $\mathcal{M}$}
\label{appToM}
The likelihood ratio defined in Eq.~\eqref{qrho} can be generalized to any matrix size, which includes $\mathcal{M}$, a $5\times4$ matrix.
In that case, the denominator is the profile likelihood for the model of the most general $5\times4$ matrix while the numerator uses the model of a general rank 1 matrix, which we compose as the tensor product of two vectors: $(\mu_{\gamma\gamma}, \mu_{\mathrm{WW}}, \mu_{\mathrm{ZZ}}, \mu_{\tau\tau},\mu_{\mathrm{bb}})$ and $(1, \lambda_{\mathrm{VBF}}, \lambda_\mathrm{VH}, \lambda_\mathrm{ttH})$, where
$
\mu_{j}=\mu_{\mathrm{ggH},j}
$
is the signal strength for gluon fusion production and decay mode $j$, and 
$
\lambda_i= \mu_{i,j} / \mu_{\mathrm{ggH},j} 
$
are the relative scaling factors between the signal strength in production mode $i$ and that for gluon fusion.
Note that, as expected, the rank 1 model entails that the $\lambda_i$ are the same for all decay modes.

The matrix under each of the alternative hypotheses is:
\begin{enumerate}

\item A general rank 1 matrix with eight parameters $\mu_{j}$, $\lambda_{\mathrm{VBF}}$, $\lambda_{\mathrm{VH}}$, and $\lambda_{\mathrm{ttH}}$:
\[
\small
\begin{array}{r|ccccc}
& \mathrm{H}\rightarrow \gamma\gamma & \mathrm{H}\rightarrow \mathrm{WW} & \mathrm{H}\rightarrow \mathrm{ZZ} & \mathrm{H}\rightarrow \tau\tau & \mathrm{H}\rightarrow \mathrm{bb}\\
 \hline
\mathrm{ggH} & \mu_{\gamma\gamma} & \mu_{\mathrm{WW}} & \mu_{\mathrm{ZZ}} & \mu_{\tau\tau} & \mu_{\mathrm{bb}}\\[3pt]
\mathrm{VBF} & \lambda_{\mathrm{VBF}}\cdot\mu_{\gamma\gamma} & \lambda_{\mathrm{VBF}}\cdot\mu_{\mathrm{WW}} & \lambda_{\mathrm{VBF}}\cdot\mu_{\mathrm{ZZ}} & \lambda_{\mathrm{VBF}}\cdot\mu_{\tau\tau} & \lambda_{\mathrm{VBF}}\cdot\mu_{\mathrm{bb}} \\[3pt]
\mathrm{VH} & \lambda_{\mathrm{VH}}\cdot\mu_{\gamma\gamma} & \lambda_{\mathrm{VH}}\cdot\mu_{\mathrm{WW}} & \lambda_{\mathrm{VH}}\cdot\mu_{\mathrm{ZZ}} & \lambda_{\mathrm{VH}}\cdot\mu_{\tau\tau} & \lambda_{\mathrm{VH}}\cdot\mu_{\mathrm{bb}}\\[3pt]
\mathrm{ttH} & \lambda_{\mathrm{ttH}}\cdot\mu_{\gamma\gamma} & \lambda_{\mathrm{ttH}}\cdot\mu_{\mathrm{WW}} & \lambda_{\mathrm{ttH}}\cdot\mu_{\mathrm{ZZ}} & \lambda_{\mathrm{ttH}}\cdot\mu_{\tau\tau} & \lambda_{\mathrm{ttH}}\cdot\mu_{\mathrm{bb}}
\end{array}\]

\item The most general 5$\times$4 matrix with twenty parameters $\mu_{j}$, $\lambda^{j}_{\mathrm{VBF}}$, $\lambda^{j}_\mathrm{VH}$, and $\lambda^{j}_\mathrm{ttH}$:
\[
\small
\begin{array}{r|ccccc}
& \mathrm{H}\rightarrow \gamma\gamma & \mathrm{H}\rightarrow \mathrm{WW} & \mathrm{H}\rightarrow \mathrm{ZZ} & \mathrm{H}\rightarrow \tau\tau & \mathrm{H}\rightarrow \mathrm{bb}\\
 \hline
\mathrm{ggH} & \mu_{\gamma\gamma}  & \mu_{\mathrm{WW}} & \mu_{\mathrm{ZZ}} & \mu_{\tau\tau} & \mu_{\mathrm{bb}}\\[4pt]
\mathrm{VBF} & \lambda^{\gamma\gamma}_{\mathrm{VBF}}\cdot\mu_{\gamma\gamma} & \lambda^{WW}_{\mathrm{VBF}}\cdot\mu_{\mathrm{WW}} & \lambda^{\mathrm{ZZ}}_{\mathrm{VBF}}\cdot\mu_{\mathrm{ZZ}} & \lambda^{\tau\tau}_{\mathrm{VBF}}\cdot\mu_{\tau\tau} & \lambda^{bb}_{\mathrm{VBF}}\cdot\mu_{\mathrm{bb}} \\[5pt]
\mathrm{VH} & \lambda^{\gamma\gamma}_\mathrm{VH}\cdot\mu_{\gamma\gamma} & \lambda^{WW}_\mathrm{VH}\cdot\mu_{\mathrm{WW}} & \lambda^{\mathrm{ZZ}}_\mathrm{VH}\cdot\mu_{\mathrm{ZZ}} & \lambda^{\tau\tau}_\mathrm{VH}\cdot\mu_{\tau\tau} & \lambda^{bb}_\mathrm{VH}\cdot\mu_{\mathrm{bb}}\\[5pt]
\mathrm{ttH} & \lambda^{\gamma\gamma}_\mathrm{ttH}\cdot\mu_{\gamma\gamma} & \lambda^{WW}_\mathrm{ttH}\cdot\mu_{\mathrm{WW}} & \lambda^{\mathrm{ZZ}}_\mathrm{ttH}\cdot\mu_{\mathrm{ZZ}} & \lambda^{\tau\tau}_\mathrm{ttH}\cdot\mu_{\tau\tau} & \lambda^{bb}_\mathrm{ttH}\cdot\mu_{\mathrm{bb}}
\end{array}\]

\end{enumerate}

If the matrix has rank 1, the rows in these two models are the same, i.e.\  $\lambda_{\mathrm{VBF}}\equiv\lambda^{j}_{\mathrm{VBF}}$,
$\lambda_\mathrm{VH}\equiv\lambda^{j}_\mathrm{VH}$, 
and $\lambda_{\mathrm{ttH}}\equiv\lambda^{j}_\mathrm{ttH}$.
The parameters of interest are $\lambda_{\mathrm{VBF}}$ ($\lambda^{j}_{\mathrm{VBF}}$), $\lambda_\mathrm{VH}$ ($\lambda^{j}_{\mathrm{VH}}$), and $\lambda_\mathrm{ttH}$ ($\lambda^{j}_{\mathrm{ttH}}$), while the signal strength parameters are profiled as they have no bearing on the rank of the matrix.

The test statistic $q_{\lambda}$ is defined using ratio between the profile likelihood of the two aforementioned models:
\begin{equation}
\label{qratio}
q_{\lambda}=-2 \ln \frac{L(\mathrm{data}|\lambda^{j}_{\mathrm{VBF}}=\hat{\lambda}_{\mathrm{VBF}}, \ \lambda^{j}_\mathrm{VH}=\hat{\lambda}_\mathrm{VH}, \  \lambda^{j}_\mathrm{ttH}=\hat{\lambda}_{\mathrm{ttH}})}{L(\mathrm{data}|\hat{\lambda}^{j}_{\mathrm{VBF}}, \hat{\lambda}^{j}_\mathrm{VH}, \hat{\lambda}^{j}_\mathrm{ttH})}.
\end{equation}
The numerator is the profile likelihood of data assuming the most general rank 1 matrix and the denominator is the profile likelihood of data assuming the most general $5\times4$ matrix. 

As noted above, if there is only one Higgs boson the $\lambda_i$ do not depend on the decay mode and the value of the $q_{\lambda}$ is not very large.
However, if the rank is not equal to 1, the most general $5\times4$ matrix model will fit the data better than the general rank 1 matrix model and the value of $q_{\lambda}$ will be large. 

When dealing with more general parameters than the double ratio $\rho$, the expected distribution of the test statistic is generated using a Monte Carlo simulation.
The p-value for the observation is evaluated from the expected test statistic distribution under the hypothesis of the SM Higgs boson. 

\section{Numerical examples}
\label{numExample}
For ease of comparison, and even if more recent data are available~\cite{CMSHbbLegacyRun1,CMSHwwLegacyRun1,CMSHzzLegacyRun1,CMSHttLegacyRun1,CMSHggLegacyRun1,CMSHzg2013,ATLASHzg2014,ATLASInvisible2014,CMSHinvLegacyRun1,CMSttHLegacyRun1,ATLASHmm2014,ATLASHggFiducial2014,ATLASHzzFiducial2014,ATLASHzzCouplings2014,ATLASHggCouplings2014}, the following numerical examples are carried out using the same signal strength matrix presented in Ref.~\cite{Grossman}, namely:
\begin{equation}
\small
\label{numbers}
\begin{array}{r|ccccc}
 & \mathrm{H}\rightarrow \gamma\gamma & \mathrm{H}\rightarrow \mathrm{WW} & \mathrm{H}\rightarrow \mathrm{ZZ} & \mathrm{H}\rightarrow \tau\tau & \mathrm{H}\rightarrow \mathrm{bb} \\
 \hline
\mathrm{ggH} & 1.6 \pm 0.35 & 0.8 \pm 0.3 & 1.0 \pm 0.3 & 1.2 \pm 0.8 & - \\[3pt]
\mathrm{VBF} & 2.1 \pm 0.9 & -0.2 \pm 0.6 & - & 0.3 \pm 0.7 & - \\[3pt]
\mathrm{VH} & 1.9 \pm 2.6 & -0.3 \pm 2.1 & - & 1.0 \pm 1.8 & 0.8 \pm 0.6  \\[3pt] 
\mathrm{ttH} & - & - & - & - & < 3.8
\end{array}
\end{equation}

In the following sub-sections we will consider, in turn, sub-matrices of Eq.~\eqref{numbers}, building up to the analysis of all the information it provides.
While any correlated uncertainties between the matrix elements have not been taken into account, experiments can and should take them into account when using this method.

\subsection{$(\mathrm{ggH, VBF}) \times (\gamma\gamma, \mathrm{WW}, \tau\tau)$}
\label{section2x3}
As a first example, consider the $2\times3$ sub-matrix spanned by $(\mathrm{ggH, VBF}) \times (\gamma\gamma, \mathrm{WW}, \tau\tau)$ with signal strengths taken from Eq.~\eqref{numbers} above.
As an approximation, each signal strength is modeled by a Gaussian distribution using the likelihood function for a counting experiment with large event yields.
The counting experiment is set up as follows: the expected number of background events $\mathrm{B_{exp}}$ is fixed to $1000$ and the number of observed events $\mathrm{N_{obs}}$ and the expected number of signal events $\mathrm{S_{exp}}$ are chosen to reproduce the observed signal strength $\mu_{i,j}$ and its uncertainty. Using this approach, the observed value of the test statistic is $q_{\lambda}^{\mathrm{obs}}=4.02$.

\begin{figure}[h!]
\begin{minipage}{\columnwidth}
\centering
\includegraphics[width=0.80\textwidth]{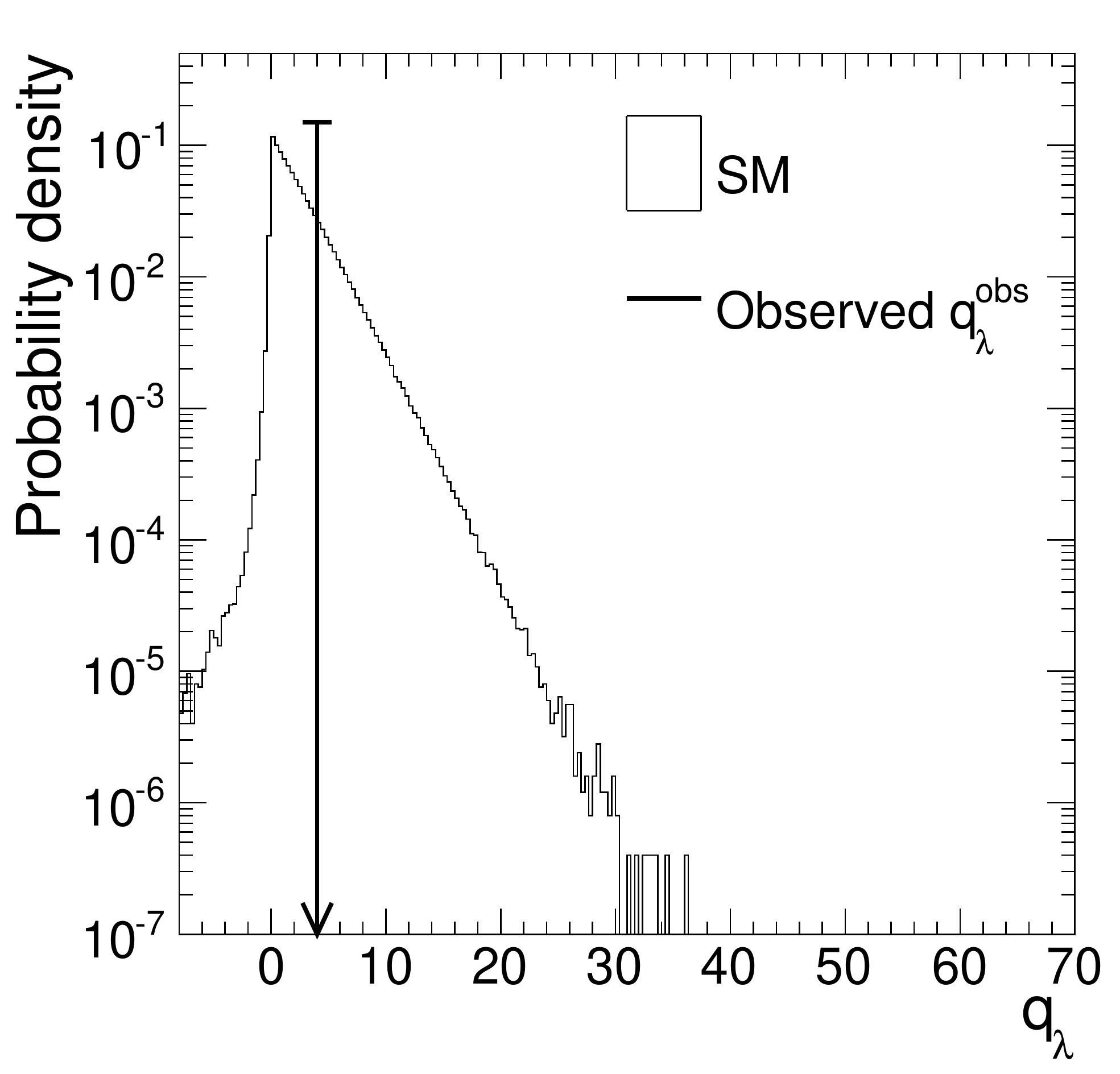}
\end{minipage}
\caption{%
The distribution of the test statistic,
generated using the $2\times3$ sub-matrix of the Eq.~\eqref{numbers},
spanned by $(\mathrm{ggH, VBF}) \times (\gamma\gamma, \mathrm{WW}, \tau\tau)$ and assuming the SM Higgs boson hypothesis.
The observed value of is $q_{\lambda}^{\mathrm{obs}}=4.02$ and,
under the SM Higgs boson hypothesis,
the fraction of pseudo-experiments for which
$q_{\lambda}\geq 4.02$ is 0.2090.
}
\label{2x3}
\end{figure}

Pseudo-experiments are used to estimate the expected distribution of the test statistic under the hypothesis of a single particle.
The pseudo-data are generated according to the expectation for the SM Higgs boson hypothesis, where $\mu_{j}=\lambda_\mathrm{VBF}=\lambda^j_\mathrm{VBF}=1$. 
The expected test statistic distribution is shown in Fig.~\ref{2x3}.
Small negative values of $q_{\lambda}$ are expected from the limited numerical precision when evaluating $q_{\lambda}$, when $q_{\lambda} \sim 0$.
On the other hand, large negative values are due to the failure to fit the pseudo-data with the more general model (in the denominator of $q_{\lambda}$) while the rank 1 model fit converges.
These are rare cases, present in this particular scenario that has large input uncertainties.
This is in contrast to the scenario considered in the following sub-section, where these large negative values are absent, as expected from the reduced uncertainties.

The p-value for the observed outcome can be derived using the distribution of the test statistic in pseudo-data shown in Fig.~\ref{2x3} by counting how many outcomes have $q_{\mathrm{\lambda}} \geq q^{\mathrm{obs}}_{\mathrm{\lambda}}$: 
\[
\begin{split}
p_{\mathrm{SM}} =& P(q_{\mathrm{\lambda}} \geq  q_{\mathrm{\lambda}}^{\mathrm{obs}} | \mathrm{\mu_j} \equiv 1, \mathrm{\lambda^j_i} \equiv 1)\\
 =&  0.2090.
\end{split}
\]
This means that if the matrix has rank 1 we would expect an outcome as extreme, or more extreme,
than the observation in $(20.90 \pm 0.03 ) \%$ of the cases, where the uncertainty quoted is purely statistical and is calculated from the 68.2\% Clopper-Pearson interval~\cite{ClopperPearson}.
The p-value can be translated to into a z score and for that conversion we use the one-tail convention.
The z score for the observed p-value above is $0.81$ 
standard deviations ($\sigma$) and represents a low significance against the rank 1 hypothesis.
This result is in contrast with the findings of Ref.~\cite{Grossman}, namely $0.4\sigma$.
However, both results are compatible with the null hypothesis.

As would be expected, there is no change in the results when extending the matrix to include also the $\mathrm{ZZ}$ information from Eq.~\eqref{numbers}.
The reason is that having one single element in any given row or column makes no statement about the rank-related parameters, $\lambda_{i}$.

\subsection{Luminosity extrapolation}
\label{luminosityExtra}

\begin{figure}
\begin{minipage}{\columnwidth}
\centering
\includegraphics[width=0.80\textwidth]{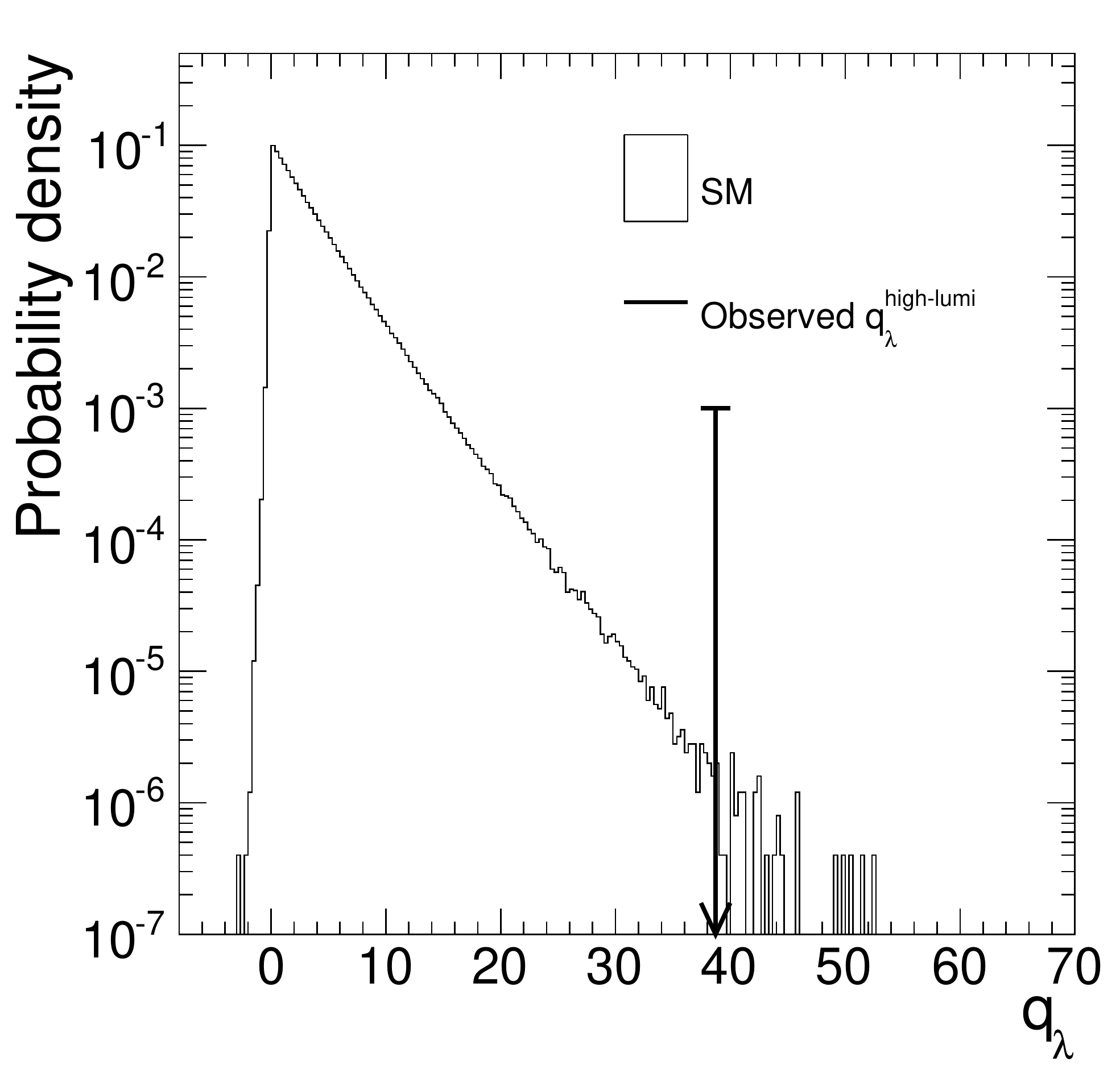}
\end{minipage}
\caption{
The distribution of the test statistic, generated using the $2\times3$ sub-matrix of the Eq.~\eqref{numbers}, spanned by $(\mathrm{ggH, VBF}) \times (\gamma\gamma, \mathrm{WW}, \tau\tau)$ and assuming the SM Higgs boson hypothesis.
The signal strength uncertainties were scaled with $1/\sqrt{L}=1/\sqrt{10}$.
The observed value of $q_{\lambda}$ is $q_{\lambda}^{\mathrm{high-lumi}} = 38.7$ and,
under the SM Higgs boson hypothesis, the fraction of pseudo-experiments for which $q_{\lambda} \geq 38.7$ is
$(1.6 \pm 0.3)\times 10^{-5}$.
}

\label{2x3ERR}
\end{figure}

For a different type of numerical example, let us examine the same $2\times3$ sub-matrix as above, but scaling the signal strength uncertainties with $1/\sqrt{L}=1/\sqrt{10}$, a naive way of simulating a 10-fold increase in the amount of data, and also done in Ref.~\cite{Grossman}.
Because of the decreased uncertainties, the observed value of the test statistic would become $q_{\lambda}^{\mathrm{high-lumi}}=38.7$.
The expected test statistic distribution can be seen in Fig.~\ref{2x3ERR} and the p-value is found to be: 
\[
\begin{split}
p_{\mathrm{SM, high-lumi}} =& P(q_{\mathrm{\lambda}} \geq  q_{\mathrm{\lambda}}^{\mathrm{high-lumi}} | \mathrm{\mu_j} \equiv 1, 
\mathrm{\lambda^j_i} \equiv 1) \\
=& ( 1.6 \pm 0.3) \times 10^{-5},
\end{split}
\]
which would correspond to a $4.15 \pm 0.04 \sigma$ significance for rank larger than unity, to be compared with $3.9\sigma$ estimated in Ref.~\cite{Grossman}.

From the comparison of the two test statistic distributions in Figs.~\ref{2x3} and \ref{2x3ERR}, one notes the effect that the reduction of the uncertainties on the signal strengths has in decreasing the number of outcomes with $q_{\lambda}<0$, as previously described. 

\subsection{$(\mathrm{ggH, VBF, VH}) \times (\gamma\gamma, \mathrm{WW}, \tau\tau)$}
\label{largestMatrix}

Let us now turn to the $3\times3$ sub-matrix in Eq.~\eqref{numbers} that is spanned by $(\mathrm{ggH, VBF, VH}) \times (\gamma\gamma, \mathrm{WW}, \tau\tau)$.

\begin{figure}[h!]
\begin{minipage}{\columnwidth}
\centering
\includegraphics[width=0.80\textwidth]{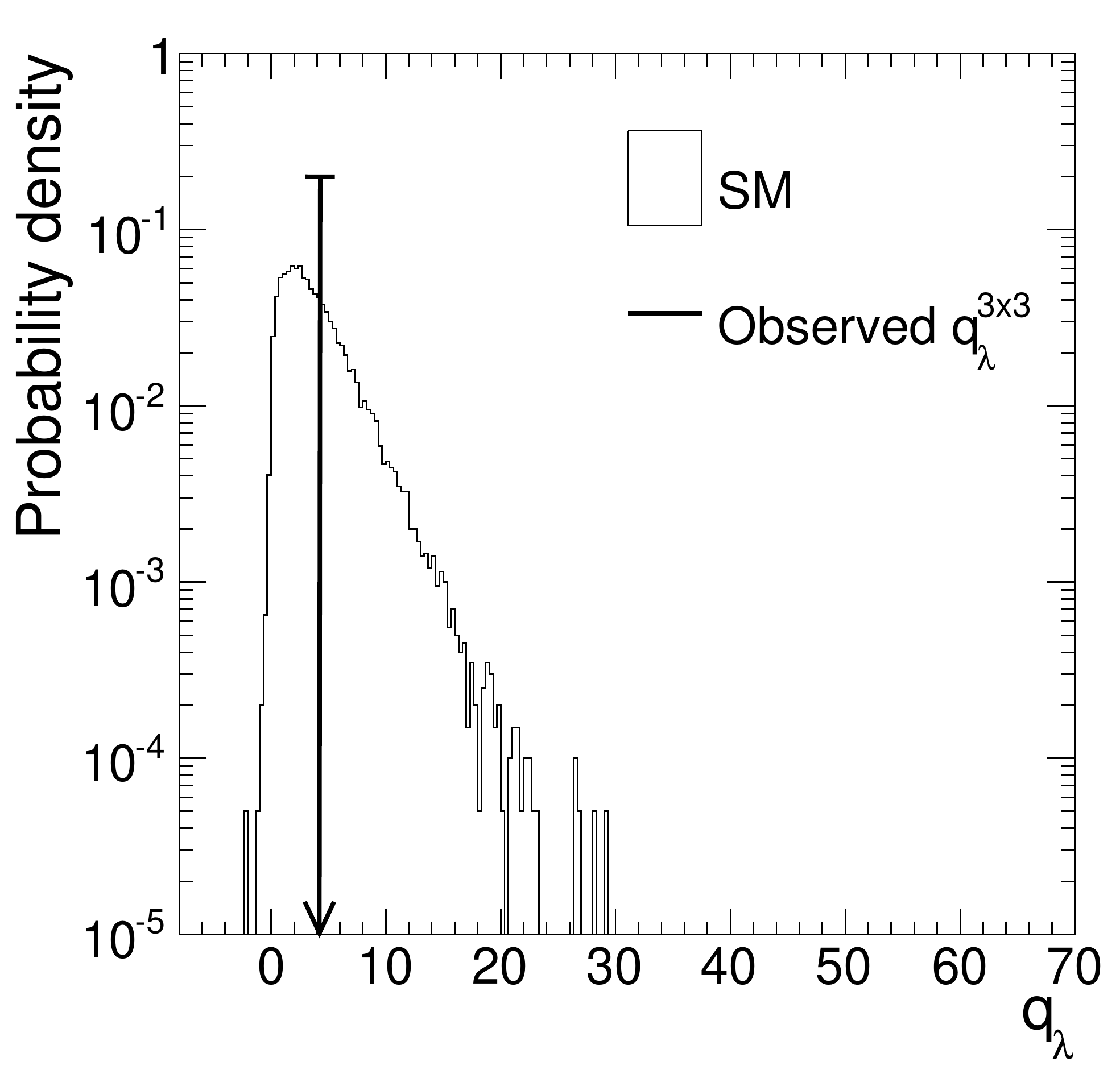}
\end{minipage}
\caption{%
The distribution of the test statistic,
generated using the $3\times3$ sub-matrix of the Eq.~\eqref{numbers},
spanned by $(\mathrm{ggH, VBF, VH}) \times (\gamma\gamma, \mathrm{WW}, \tau\tau)$ and assuming the SM Higgs boson hypothesis.  The observed value of $q_{\lambda}$ is $ q_{\lambda}^{\mathrm{3\times3}} = 4.28$ and, under the SM Higgs boson hypothesis, the fraction of pseudo-experiments for which $q_{\lambda} \geq 4.28$ is $0.35$.}
\label{3x3}
\end{figure}

The observed value of the test statistic is found to be $q_{\lambda}^{\mathrm{3\times3}}=4.28$.
The expected test statistic distribution can be seen in Fig.~\ref{3x3} and the p-value is: 
\[
\begin{split}
p_{\mathrm{SM, 3x3}} =& P(q_{\mathrm{\lambda}} \geq  q_{\mathrm{\lambda}}^{\mathrm{3\times3}} | \mathrm{\mu_j} \equiv 1, 
\mathrm{\lambda^j_i} \equiv 1) \\
=& 0.35,
\end{split}
\]
which corresponds to a $0.4\sigma $ significance for rank larger than unity.

\subsection{$(\mathrm{ggH, VBF, VH}) \times (\gamma\gamma, \mathrm{WW}, \tau\tau)$ with removed elements}
\label{removedElements}
To demonstrate the way in which the proposed technique deals with incomplete data, two different tests with removed elements are presented.

\paragraph{Two elements removed}
In a first test, the two elements with large uncertainties, namely $\mu_{\mathrm{VBF},\gamma\gamma}$ and $\mu_{\mathrm{VH,WW}}$, have been removed.

\begin{figure}[h!]
\begin{minipage}{\columnwidth}
\centering
\includegraphics[width=0.80\textwidth]{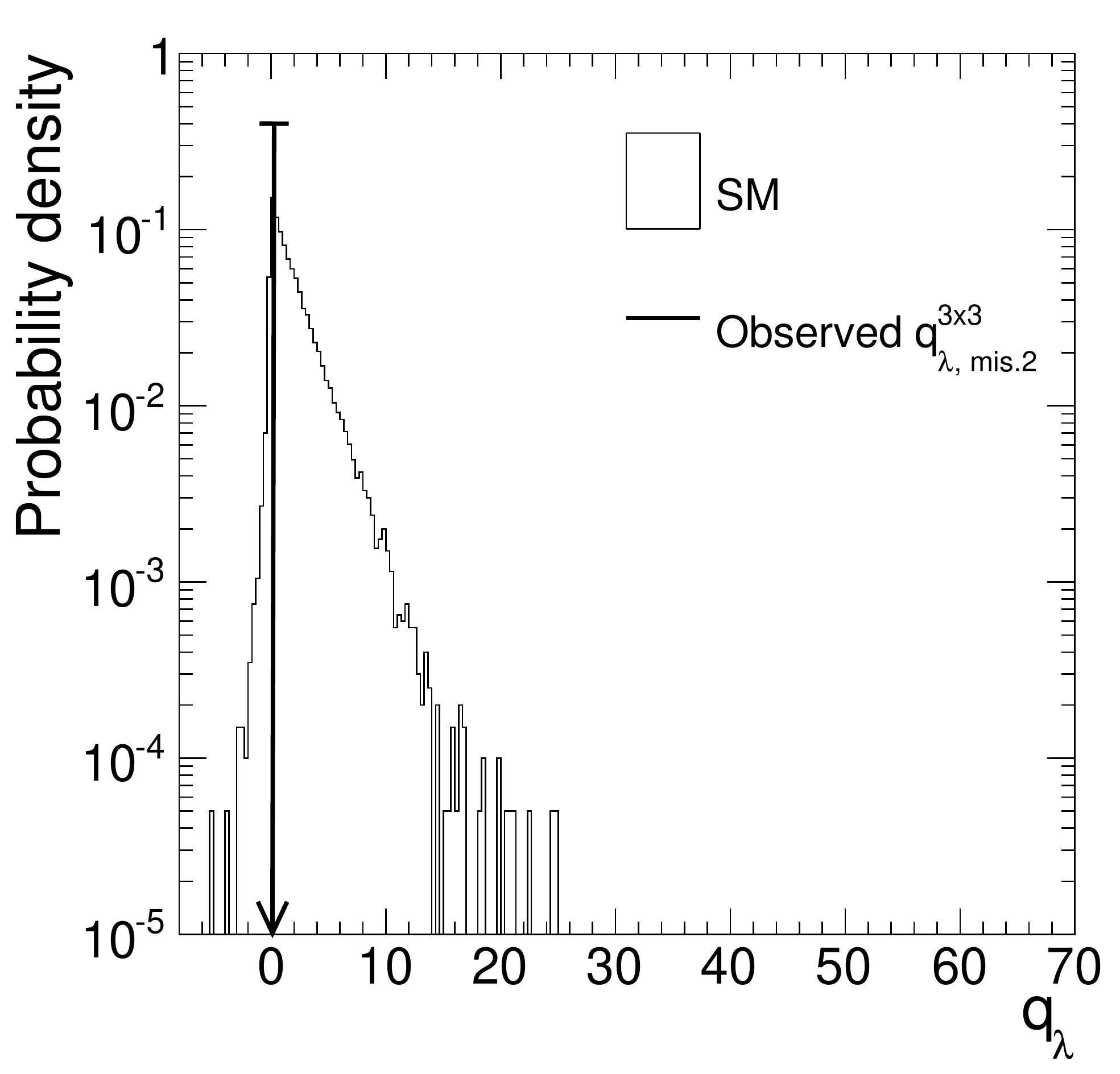}
\end{minipage}
\caption{%
The distribution of the test statistic,
generated using the $3\times3$ sub-matrix of the Eq.~\eqref{numbers},
spanned by $(\mathrm{ggH, VBF, VH}) \times (\gamma\gamma, \mathrm{WW}, \tau\tau)$, with the signal strengths $\mu_{\mathrm{VBF},\gamma\gamma}$ and $\mu_{\mathrm{VH,WW}}$ removed, and assuming the SM Higgs boson hypothesis.
The observed value of $q_{\lambda}$ is $ q_{\lambda, \mathrm{mis. 2}}^{\mathrm{3\times3}} = 0.28$ and, under the SM Higgs boson hypothesis, the fraction of pseudo-experiments for which $q_{\lambda} \geq 0.28$ is $0.81$.}
\label{3x3MIS2}
\end{figure}

The observed value of the test statistic is found to be $q_{\lambda, \mathrm{mis. 2}}^{\mathrm{3\times3}}=0.28$ and the p-value, defined from the test statistic distribution in Fig. \ref{3x3MIS2}, is
\[\begin{split}
p_\mathrm{SM, 3\times3, mis. 2} =& P(q_{\mathrm{\lambda}} \geq  q_{\mathrm{\lambda, mis. 2}}^{\mathrm{3\times3}} | \mathrm{\mu_j} \equiv 1, 
\mathrm{\lambda^j_i} \equiv 1) \\
=& 0.81, 
\end{split}\]
which corresponds to a $-0.86\sigma$
significance for rank larger than unity, where the negative sign is a reflection of $q_{\mathrm{\lambda, mis. 2}}^{\mathrm{3\times3}}$ being lower than the median of the SM distribution.

\paragraph{Three elements removed}

In a second test, we remove the $\mu_{\mathrm{VH},\gamma\gamma}$, $\mu_{\mathrm{VBF,WW}}$, and $\mu_{\mathrm{ggH},\tau\tau}$ elements from the $(\mathrm{ggH, VBF, VH}) \times (\gamma\gamma, \mathrm{WW}, \tau\tau)$ matrix.

In this case, a signal strength measurement $\mu_{\tau\tau}$ is absent, implying that the parameterization for the most general $5\times4$ matrix has more free parameters than there are measurements.
This presents a technical difficulty for the numerical minimization used in calculating the profile likelihood.
Nevertheless, and without loss of generality, the missing $\mu_{j}$ can be set to unity in the denominator of $q_{\lambda}$,  Eq.~\eqref{qratio}.

\begin{figure}[h!]
\begin{minipage}{\columnwidth}
\centering
\includegraphics[width=0.80\textwidth]{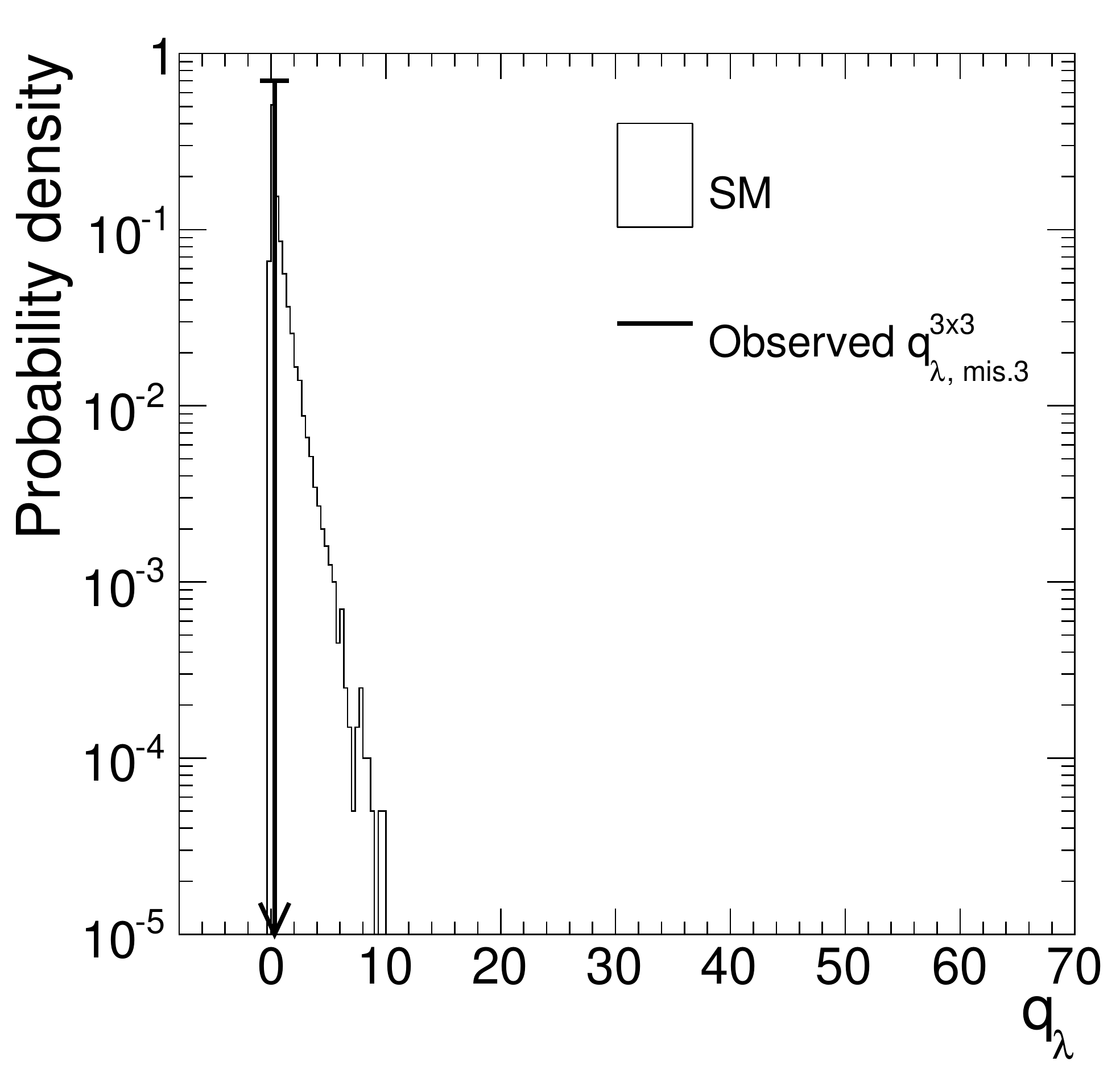}
\end{minipage}
\caption{%
The distribution of the test statistic,
generated using the $3\times3$ sub-matrix of the Eq.~\eqref{numbers},
spanned by $(\mathrm{ggH, VBF, VH}) \times (\gamma\gamma, \mathrm{WW}, \tau\tau)$, after removing three elements, and assuming the SM Higgs boson hypothesis. The observed value of $q_{\lambda}$ is $ q_{\lambda, \mathrm{mis. 3}}^{\mathrm{3\times3}} = 0.31$ and, under the SM Higgs boson hypothesis, the fraction of pseudo-experiments for which $q_{\lambda} \geq 0.31$ is $0.44$.}
\label{3x3MISAD}
\end{figure}

The observed value of the test statistic is found to be $q_{\lambda, \mathrm{mis. 3}}^{\mathrm{3\times3}}=0.31$.
The test statistic distribution under the SM Higgs hypothesis can be seen in Fig. \ref{3x3MISAD}
 and the p-value is 
\[
\begin{split}
p_{\mathrm{SM, 3\times3, mis. 3}} =& P(q_{\mathrm{\lambda}} \geq  q_{\mathrm{\lambda, mis. 3}}^{\mathrm{3\times3}} | \mathrm{\mu_j} \equiv 1, 
\mathrm{\lambda^j_i} \equiv 1) \\
=& 0.44,
\end{split}
\]
which corresponds to a $0.15\sigma$ 
significance for rank larger than unity.

\section*{Conclusion}
\label{conclusion}
We have developed a method that can be used to test for the presence of multiple Higgs bosons.
The method can be directly used by the ATLAS and CMS collaborations as it builds upon the profile likelihood techniques already in use for Higgs measurements at the LHC.
The main features of method are that it can test for (in)dependence in arbitrarily-sized linear systems in the presence of uncertainties and of missing elements of data.

\bibliographystyle{custom}
\bibliography{reference}   

\end{document}